\documentclass[12pt]{article}
\usepackage{epsfig}
\begin{document}
\baselineskip=16pt
\begin{titlepage}
\setcounter{page}{0}
\begin{center}

\vspace{0.5cm}
 {\Large \bf  Cosmological Evolution of a Quintom Model of Dark Energy}\\
\vspace{10mm}

Zong-Kuan Guo\footnote{e-mail address: guozk@itp.ac.cn}$^{b}$,
Yun-Song Piao$^{c}$,
Xinmin Zhang$^{d}$
and
Yuan-Zhong Zhang$^{a,b}$ \\
\vspace{6mm} {\footnotesize{\it
  $^a$CCAST (World Lab.), P.O. Box 8730, Beijing 100080\\
  $^b$Institute of Theoretical Physics, Chinese Academy of Sciences,
      P.O. Box 2735, Beijing 100080, China\\
  $^c$Interdisciplinary Center of Theoretical Studies, Chinese Academy
      of Sciences, P.O. Box 2735, Beijing 100080, China\\
  $^d$Institute of High Energy Physics, Chinese Academy of Science,
      P.O. Box 918-4, Beijing 100039, China\\ }}

\vspace*{5mm} \normalsize
\smallskip
\medskip
\smallskip
\end{center}
\vskip0.6in
\centerline{\large\bf Abstract}
{We investigate in this paper the cosmological evolution of a dark
energy model with two scalar fields where one of the scalar has
canonical kinetic energy and another scalar has negative kinetic energy
term. For such a system with exponential potentials we find that
during the evolution of the universe the equation of state $w$
changes from $w>-1$ to $w<-1$,  which is consistent with the recent
observations. A phase-plane analysis shows that the
``phantom"-dominated scaling solution is the stable late-time attractor
of this type of models.}
\vspace*{2mm}

\begin{flushleft}
PACS number(s): 98.80.Cq, 98.80.-k
\end{flushleft}

\end{titlepage}


Scalar fields play an important role in modern cosmology. The dark
energy can be attributed to the dynamics of a scalar or multi-scalar
fields, for instance the quintessence \cite{RP, ZWS}, which convincingly
realize the present-day cosmic acceleration by using late-time attractor
solutions, in which the scalar fields mimic the perfect fluid in a wide
range of parameters. For the detail studies on the models much
attention has been drawn to the case of exponential potentials. The
exponential potentials allow the possible existence of scaling solutions
in which the scalar field energy density tracks that of the perfect fluid
(so that at late times neither field is negligible). In particular, a
phase-plane analysis of the spatially flat FRW models showed that
these solutions are the unique late-time attractors whenever they
exist~\cite{HW, ZKG}. Moreover, exponential potentials often
appear naturally in models after compactification in string/M-theory.

By fitting the recent SNe Ia data, marginal (2$\sigma$) evidence for
$w(z)<-1$ at $z<0.2 $ has been found \cite{HC}. To obtain $w<-1$,
phantom field with a negative kinetic term may be a simplest
implementing, in which the weak energy condition is violated, and can
be regarded as one of interesting possibilities describing dark
energy~\cite{RRC}. The physical background for phantom type of
matter with strongly negative pressure may be looked for in string
theory~\cite{MGK}. Phantom field may also arise from a bulk viscous
stress due to the particle production~\cite{JDB} or in higher-order
theories of gravity~\cite{MDP}, Brans-Dicke and non-minimally
coupled scalar field theories~\cite{DFT}. The cosmological models
which allow for phantom matter appear naturally in the mirage
cosmology of the braneworld scenario~\cite{KK} and in k-essence
models~\cite{COY}. In spite of the fact that the field theory of
phantom fields encounters the problem of stability which one could
try to bypass by assuming them to be effective fields~\cite{CHT,GWG},
it is nevertheless interesting to study their cosmological implication.
Recently, there are many relevant studies on phantom energy~\cite{SW}
and the primordial perturbation spectrum from various phantom
inflation models~\cite{PIA}.

Furthermore, the analysis on the properties of dark energy from the
recent observations mildly favor models with $w$ crossing $-1$ in the
near past. However, neither quintessence nor phantom can fulfill this
transition. But in an universe with a quintessence and a phantom this
case can be realized easily. This implement of dark energy, called as
quintom, has been proposed in Ref.~\cite{FWZ}, and in some cases
providing a better fit to the data than the more familiar models with
$w \geq -1$. The quintom might be able to preserve the tracking
behavior of the quintessence and needs less fine-tuned in the early
universe compared with the phantom field, furthermore predict a
interesting feature in the evolution and fate of the universe~\cite{FLPZ}.
In this paper we study the the quintom model above with an
exponential potential and investigate the existence and stability of
cosmological scaling solutions in the context of spatially homogeneous
cosmological models. Our phase-plane analysis of the spatially flat
FRW models shows that the phantom-dominated scaling solution is the
unique late-time attractor and there exists a transition from $w>-1$
to $w<-1$, then to a constant related to the slope of the phantom field
potential at late times. We will also discuss the physical consequences
of these results.


We consider a toy model which contains a negative-kinetic scalar field 
$\phi$ and a 
normal
scalar field $\sigma$ with generic exponential potential, described by the
action:
\begin{equation}
S=\int d^4x\sqrt{-g}\left(\frac{R}{2\kappa ^2}
 -\frac{1}{2}g^{\mu \nu}\partial _\mu \phi \partial _\nu \phi
 +\frac{1}{2}g^{\mu \nu}\partial _\mu \sigma \partial _\nu \sigma
 +V(\phi,\sigma)+\mathcal{L_{\textrm{m}}}\right),
\end{equation}
where $\kappa ^2\equiv 8 \pi G_N$ is the gravitational coupling and
$\mathcal{L_{\textrm{m}}}$ represents the Lagrangian density of matter
fields. The homogeneous fields $\phi$ and $\sigma$ in a spatially flat
FRW cosmological model can be described by a fluid with an effective
energy density $\rho$ and an effective pressure $P$ given by
\begin{eqnarray}
\rho &=& -\frac{1}{2}\dot{\phi}^2+\frac{1}{2}\dot{\sigma}^2+V(\phi,\sigma), \\
P    &=& -\frac{1}{2}\dot{\phi}^2+\frac{1}{2}\dot{\sigma}^2-V(\phi,\sigma).
\end{eqnarray}
The corresponding equation of state parameter is now given by
\begin{equation}
\label{ES}
w=\frac{-\dot{\phi}^2+\dot{\sigma}^2-2V(\phi,\sigma)}
 {-\dot{\phi}^2+\dot{\sigma}^2+2V(\phi,\sigma)}.
\end{equation}
For a model with a normal scalar field, the equation of state $w \ge 
-1$. The
toy model of a phantom energy component with a negatice kinetic 
term possesses an equation of
state $w < -1$. In our model, Eq.(\ref{ES}) implies $w \ge -1$ when
$\dot{\sigma}\ge\dot{\phi}$ and $w<-1$ when $\dot{\sigma}<\dot{\phi}$.
We assume that there is no direct coupling between the phantom field
and the normal scalar field with such a potential
\begin{equation}
V(\phi,\sigma)=V_{\phi}(\phi)+V_{\sigma}(\sigma)=V_{\phi 0}\,
 e^{-\lambda_\phi \kappa \phi}+V_{\sigma 0}\,
 e^{-\lambda_\sigma \kappa \sigma},
\end{equation}
where $\lambda_\phi$ and $\lambda_\sigma$ are two dimensionless
constants characterising the slope of the potential for $\phi$ and
$\sigma$ respectively. Further we assume $\lambda_\phi \ge 0$ and
$\lambda_\sigma \ge 0$ since we can make them positive through
$\phi \to -\phi$ and $\sigma \to -\sigma$ if some of them are
negative, respectively. The evolution equations of the fields and the
fluid for a spatially flat FRW model with Hubble parameter $H$ is
\begin{eqnarray}
\label{EE1}
\ddot{\phi}+3H\dot{\phi}-\frac{dV_{\phi}(\phi)}{d\phi} &=& 0, \\
\ddot{\sigma}+3H\dot{\sigma}+\frac{dV_{\sigma}(\sigma)}{d\sigma}
 &=& 0, \\
\dot{\rho}_\gamma+3H(\rho_\gamma+P_\gamma) &=& 0,
\label{EE2}
\end{eqnarray}
where $\rho_\gamma$ is the density of fluid with a barotropic equation
of state $P_\gamma=(\gamma-1)\rho_\gamma$, where $\gamma$ is a
constant, $0 < \gamma \le 2$, such as radiation ($\gamma=4/3$) or
dust ($\gamma=1$). The Fridemann constraint equation is
\begin{equation}
\label{FE}
H^2=\frac{\kappa ^2}{3}\left(-\frac{1}{2}\dot{\phi}^2+V_{\phi}(\phi)
 +\frac{1}{2}\dot{\sigma}^2+V_{\sigma}(\sigma)+\rho_\gamma\right).
\end{equation}
Defining five dimensionless variables
\begin{eqnarray}
x_\phi \equiv \frac{\kappa \dot{\phi _i}}{\sqrt{6}H}&,& \quad y_\phi
 \equiv \frac{\kappa \sqrt{V_\phi}}{\sqrt{3}H}, \nonumber \\
x_\sigma \equiv \frac{\kappa \dot{\sigma _i}}{\sqrt{6}H}&,& \quad y_\sigma
 \equiv \frac{\kappa \sqrt{V_\sigma}}{\sqrt{3}H}, \\
z \equiv \frac{\kappa \sqrt{\rho_\gamma}}{\sqrt{3}H}&,& \nonumber
\end{eqnarray}
the evolution equations (\ref{EE1})-(\ref{EE2}) can be rewritten as an
autonomous system:
\begin{eqnarray}
\label{AS1}
x'_\phi &=& -3x_\phi\left(1+x_{\phi}^2-x_{\sigma}^2-\frac{\gamma}{2}
 z^2\right)-\lambda_\phi \frac{\sqrt{6}}{2}y_{\phi}^2\,, \\
y'_\phi &=& 3y_\phi\left(-x_{\phi}^2+x_{\sigma}^2+\frac{\gamma}{2}
 z^2-\lambda_\phi \frac{\sqrt{6}}{6}x_\phi\right), \\
x'_\sigma&=&-3x_\sigma\left(1+x_{\phi}^2-x_{\sigma}^2-\frac{\gamma}{2}
 z^2\right)+\lambda_\sigma \frac{\sqrt{6}}{2}y_{\sigma}^2\,, \\
y'_\sigma &=& 3y_\sigma\left(-x_{\phi}^2+x_{\sigma}^2+\frac{\gamma}{2}
 z^2-\lambda_\sigma \frac{\sqrt{6}}{6}x_\sigma\right), \\
z' &=& 3z\left(-x_{\phi}^2+x_{\sigma}^2+\frac{\gamma}{2}
 z^2-\frac{\gamma}{2}\right),
\label{AS2}
\end{eqnarray}
where a prime denotes a derivative with respect to the logarithm of the
scale factor, $N \equiv \ln a$, and the Fridemann constraint
equation (\ref{FE}) becomes
\begin{equation}
\label{AS3}
-x_{\phi}^2+y_\phi^2+x_{\sigma}^2+y_\sigma^2+z^2=1.
\end{equation}
We will restrict our discussion of the existence and stability of critical
points to expanding universes with $H>0$. Critical points correspond
to fixed points where $x'_\phi=0$, $y'_\phi=0$, $x'_\sigma=0$,
$y'_\sigma=0$ and $z'=0$, and there are self-similar solutions with
\begin{equation}
\frac{\dot{H}}{H^2}=3x_{\phi}^2-3x_{\sigma}^2-\frac{3\gamma}{2}z^2.
\end{equation}
This corresponds to an expanding universe with a scale factor $a(t)$
given by $a\propto t^p$, where
\begin{equation}
p=\frac{2}{-6x_{\phi}^2+6x_{\sigma}^2+3\gamma z^2}\,.
\end{equation}
The system (\ref{AS1})-(\ref{AS2}) has at most one two-dimensional
hyperbola $K$ embedded in five-dimensional phase-space corresponding
to kinetic-dominated solutions, a fixed point $P$ which is a
phantom-dominated solution, a fixed point $S$ which is a
scalar-dominated solution, a fixed point $F$ which is a fluid-dominated
solution, and a fixed point $T$ which is a fluid-scalar-dominated
solution listed in Table 1.

\begin{table}
\begin{tabular}{c c c c c c c} \hline
Label & $x_\phi$ & $y_\phi$ & $x_\sigma$ & $y_\sigma$ & z
 & Stability \\ \hline
$K$ & $x_{\sigma}^2-x_{\phi}^2=1$ & 0 & & 0 & 0 & unstable \\
$P$ & $-\frac{\lambda_\phi}{\sqrt{6}}$
 & $\sqrt{ (1+\frac{\lambda_\phi^2}{6})}$ & 0 & 0 & 0 & stable  \\
$S$ & 0 & 0 & $\frac{\lambda_\sigma}{\sqrt{6}}$ &
 $\sqrt{ (1-\frac{\lambda_\sigma^2}{6})}$ & 0 & unstable  \\
$F$ & 0 & 0 & 0 & 0 & 1 & unstable \\
$T$ & 0 & 0 & $\frac{3\gamma}{\sqrt{6}\lambda_\sigma}$
 & $\sqrt{\frac{3\gamma(2-\gamma)}{2\lambda_\sigma^2}}$
 & $\sqrt{1-\frac{3\gamma}{\lambda_\sigma^2}}$ & unstable \\ \hline
\end{tabular}
\caption{The properties of the critical points in a spatially flat FRW
universe containing a phantom field and a normal scalar field with
exponential potentials.}
\end{table}

In order to study the stability of the critical points, using the Friedmann
constraint equation (\ref{AS3}) we first reduce Eqs.(\ref{AS1})-(\ref{AS2})
to four independent equations. Substituting linear perturbations
$x_\phi \to x_\phi+\delta x_\phi$, $y_\phi \to y_\phi+\delta y_\phi$,
$x_\sigma \to x_\sigma+\delta x_\sigma$ and
$y_\sigma \to y_\sigma+\delta y_\sigma$ about the critical points into
the four independent equations, to first-order in the perturbations, gives
the evolution equations of the linear perturbations, which yield four
eigenvalues $m_i$. Stability requires the real part of all eigenvalues
to be negative.

$K$: These kinetic-dominated solutions always exist for any form of
the potential, which are equivalent to stiff-fluid dominated evolution
with $a\propto t^{1/3}$ irrespective of the nature of the potential.
The linearization of system (\ref{AS1})-(\ref{AS3}) about these fixed
points yields four eigenvalues
\begin{displaymath}
m_1=0, \quad m_2=3, \quad m_3=3(2-\gamma),
 \quad m_4=3(1 \pm \frac{\lambda_\sigma}{\sqrt{6}}),
\end{displaymath}
where we use upper/lower signs to denote the two distinct cases of
$x_\sigma = \mp \sqrt{1+x_\phi^2}$. Thus the kinetic-dominated
solutions are always unstable.

$P$: The phantom-dominated solution exist for any $\lambda_\phi$
and $\lambda_\sigma$. The power-law exponent,
$p=-2/ \lambda_\phi^2$, depends on the slope of the potential
$V_\phi$. The equation of state becomes
$w=-1-\lambda_\phi^2/3$. The linearization of system
(\ref{AS1})-(\ref{AS3}) about this fixed point yields four eigenvalues
\begin{displaymath}
m_1=-\frac{\lambda_\phi^2}{2}, \quad m_2=m_3=-\frac{1}{2}(6
 +\lambda_\phi^2), \quad m_4=-(3\gamma+\lambda_\phi^2),
\end{displaymath}
which indicate that the solution is stable.

$S$: The scalar-dominated solution exist for $\lambda_\sigma^2<6$.
The power-law exponent, $p=2/ \lambda_\sigma^2$, depends on the
slope of the potential $V_\sigma$. The equation of state becomes
$w=-1+\lambda_\sigma^2/3$. The linearization of system
(\ref{AS1})-(\ref{AS3}) about this critical point yields four eigenvalues
\begin{displaymath}
m_1=\frac{\lambda_\sigma^2}{2}, \quad m_2=m_3=\frac{1}{2}
 (\lambda_\sigma^2-6), \quad m_4=\lambda_\sigma^2-3\gamma,
\end{displaymath}
which indicate that the solution is unstable.

$F$: The fluid-dominated solution exists for any form of the
potential, corresponding to a power-law solution with $p=2/3\gamma$.
The linearization of system (\ref{AS1})-(\ref{AS3}) about this critical
point yields four eigenvalues
\begin{displaymath}
m_1=m_2=\frac{3\gamma}{2}, \quad m_3=m_4=\frac{3\gamma}{2}-3,
\end{displaymath}
which indicate that the solution is unstable.

$T$: The scalar-fluid-dominated solution exist for a potential with
$\lambda_\sigma^2 > 3\gamma$. The power-law exponent,
$p=2/3\gamma$, is identical to that of the fluid-dominated solution,
depends only on the barotropic index $\gamma$ and is independent of
the slope $\lambda_\sigma$ of the potential $V_\sigma$. The
linearization of system (\ref{AS1})-(\ref{AS3}) about the fixed point
yields four eigenvalues
\begin{eqnarray}
m_1 &=& \frac{3\gamma}{2}, \nonumber \\
m_2 &=& \frac{3\gamma}{2}-3, \nonumber \\
m_3 &=& -\frac{3(2-\gamma)}{4}\left(1+\sqrt{1-\frac{8\gamma(\lambda_
 \sigma^2-3\gamma)}{\lambda_\sigma^2(2-\gamma)}}\right) , \nonumber \\
m_4 &=& -\frac{3(2-\gamma)}{4}\left(1-\sqrt{1-\frac{8\gamma(\lambda_
 \sigma^2-3\gamma)}{\lambda_\sigma^2(2-\gamma)}}\right) , \nonumber
\end{eqnarray}
which indicate that the solution is unstable.

\begin{figure}
\begin{center}
\includegraphics[scale=1.2]{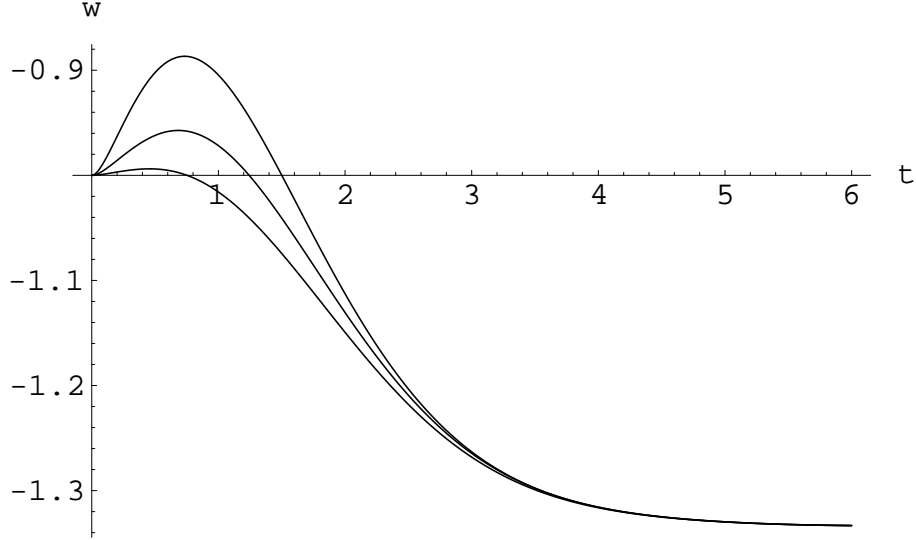}
\caption{The evolution of the effective equation of state of
the phantom and normal scalar fields with
$V(\phi,\sigma)=V_{\phi 0}\,e^{-\lambda_\phi \kappa \phi}
+V_{\sigma 0}\,e^{-\lambda_\sigma \kappa \sigma}$
for the case $\lambda_{\phi}=1$.}
\end{center}
\end{figure}


The case with the evolution of the state equation parameter $w$ crossing
$-1$, a scenario of quintom has be shown to be favored mildly by the 
recent observations. In this paper we have discussed a possible realization
of it, in which both quintessence field and phantom field are introduced.
We have presented a phase-space analysis of the evolution for a spatially
flat FRW universe containing a barotropic fluid and phantom-scalar fields
with exponential potentials and shown that the phantom-dominated scaling
solution is the stable late-time attractor. The energy density of the
phantom field dominates at the late time. The reason for this behavior is 
that the energy density of the phantom field increases while those of the
normal scalar field and the barotropic fluid decreases as the universe
evolves. Our numerical studies indicate that the state equation parameter
$w$ changes from above $-1$ to below $-1$ and tends to be
$-(1+\lambda_\phi^2/3)$ as shown in Figure 1. Moreover, for various
selections of potentials, we find that the state equation parameter $w$
changes from above $-1$ to below $-1$ and tends to be $-1$ as
shown in Figure 2, and from below $-1$ to above $-1$ and tends
to be $-1$ in Figure 3. we have assumed that there is no direct coupling
between the phantom field and the normal scalar field in this paper.

\begin{figure}
\begin{center}
\includegraphics[scale=1.2]{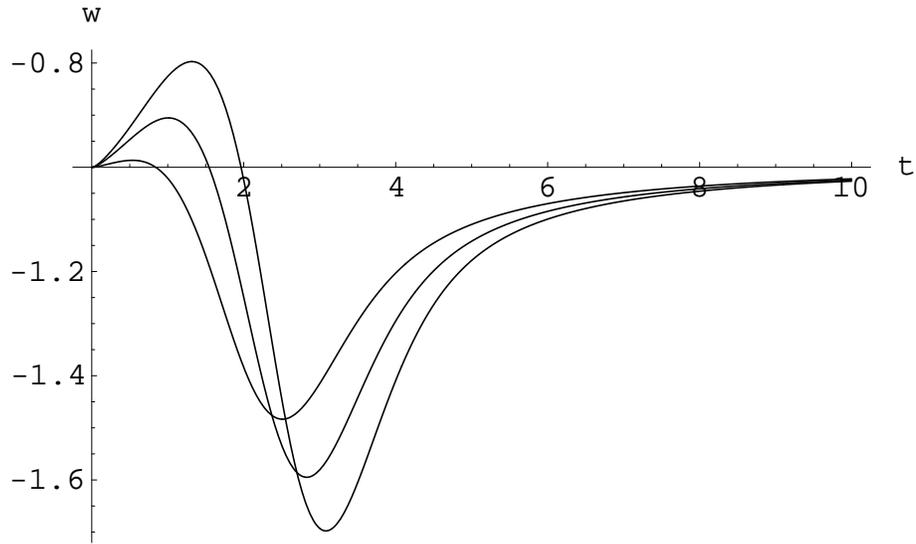}
\caption{The evolution of the effective equation of state of
the phantom and normal scalar fields with
$V(\phi,\sigma)=\frac{1}{2}m_\phi^2 \phi^2
+\frac{1}{2}m_\sigma^2 \sigma^2$.}
\end{center}
\end{figure}
\begin{figure}
\begin{center}
\includegraphics[scale=1.2]{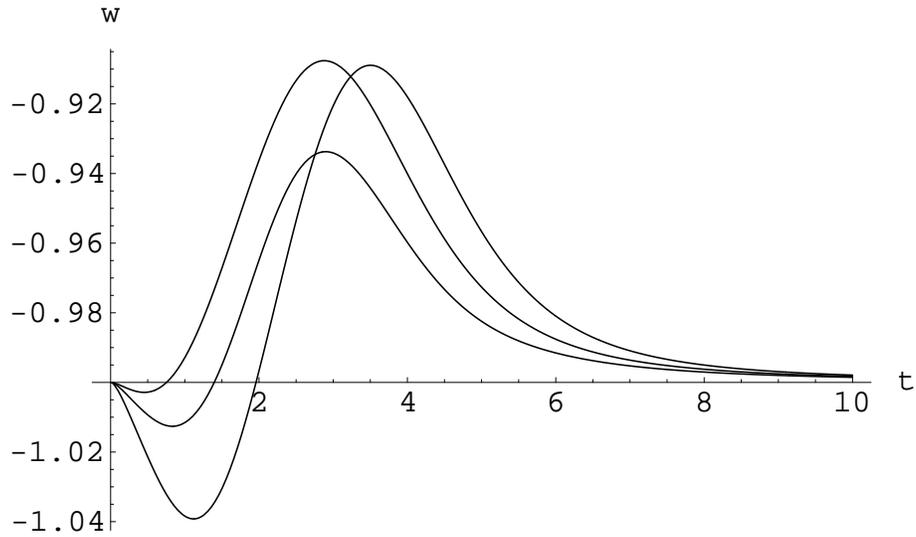}
\caption{The evolution of the effective equation of state of
the phantom and normal scalar fields with
$V(\phi,\sigma)=V_{\phi 0}\,e^{-\lambda_\phi \kappa^2 \phi^2}
+V_{\sigma 0}\,e^{-\lambda_\sigma \kappa^2 \sigma^2}$.}
\end{center}
\end{figure}

\section*{Acknowledgements}
This project was in part supported by National Basic Research Program
of China under Grant No.2003CB716300 and also by NNSFC under
Grant No.10175070.

\end{document}